\documentclass[aps,pre,preprint,amsmath,amssymb,nofootinbib]{revtex4-1}
\usepackage[percent]{overpic}

\begin{document}
\begin{flushleft}
\textit{Journal of Experimental and Theoretical Physics, 2011, Vol. 113, No. 4, pp. 605--618.}\footnote{%
\baselineskip=5pt
Original Russian Text has been published in
Zhurnal Eksperimental'noi i Teoreticheskoi Fiziki, 2011, Vol.~140, No.~4, pp.~696--711.}
\end{flushleft}
\title{Absorption of Gamma-Ray Photons\\
in a Vacuum Neutron Star Magnetosphere:\\
II. The Formation of ``Lightnings''}
\author{Ya.\ N. \surname{Istomin}}
\email{\texttt{istomin@lpi.ru}}
\author{D. N. \surname{Sob'yanin}}
\email{\texttt{sobyanin@lpi.ru}}
\affiliation{Lebedev Physical Institute, Russian Academy of Sciences,\\Leninskii pr.\ 53, Moscow, 119991 Russia}
\received{December 1, 2010}
\begin{abstract}
The absorption of a high-energy photon from the external cosmic gamma-ray background in the inner neutron star magnetosphere triggers the generation of a secondary electron-positron plasma and gives rise to a lightning---a lengthening and simultaneously expanding plasma tube. It propagates along magnetic fields lines with a velocity close to the speed of light. The high electron-positron plasma generation rate leads to dynamical screening of the longitudinal electric field that is provided not by charge separation but by electric current growth in the lightning. The lightning radius is comparable to the polar cap radius of a radio pulsar. The number of electron-positron pairs produced in the lightning in its lifetime reaches~$10^{28}$. The density of the forming plasma is comparable to or even higher than that in the polar cap regions of ordinary pulsars. This suggests that the radio emission from individual lightnings can be observed. Since the formation time of the radio emission is limited by the lightning lifetime, the possible single short radio bursts may be associated with rotating radio transients (RRATs).
\end{abstract}
\maketitle
\newpage
\section{INTRODUCTION}

This paper is a continuation of our previous paper \citep{IstominSobyanin2011a}, in which we calculated the source of electron-positron pairs for the case of a strong electric field in a vacuum neutron star magnetosphere. Interest in this subject matter is related to the appearance of new observational data on nonstationary radio sources, such as intermittent pulsars \citep{KramerEtal2006} and rotating radio transients (RRATs) \citep{McLaughlinEtal2006,McLaughlinEtal2009}. These sources, just as ordinary radio pulsars, are associated with neutron stars. In this case, the corresponding surface magnetic field is high and reaches $5\times10^{13}$~G \citep{McLaughlinEtal2006,EsamdinEtal2008}, i.e., it exceeds the Schwinger field, for example, for RRAT J1819--1458. Magnetars, which are also characterized by radio emission nonstationarity observed at low \citep{ShitovEtal2000,MalofeevEtal2005,MalofeevEtal2007,MalofeevEtal2010} and high \citep{CamiloEtal2006,CamiloEtal2007,LevinEtal2010} radio frequencies, have an even stronger magnetic field. Note that the presence of such a strong magnetic field and an electron-positron plasma changes the kinematics of quantum processes \citep{RumyantsevChistyakov2005,RumyantsevChistyakov2009}. For example, photon splitting becomes significant \citep{BaierEtal1996,Chistyakov1998}. However, it does not suppress the generation of an electron-positron plasma in a superstrong magnetic field and, consequently, the corresponding radio emission \citep{IstominSobyanin2007,IstominSobyanin2008}, although the radio emission itself can be observed erratically. In particular, there are data on the switch-off of a magnetar without any accompanying X-ray variability \citep{LevinEtal2010}. Since plasma outflows from the magnetosphere are currently believed to be responsible for the observed radio emission from neutron stars (see, e.g., the review \citep{Beskin1999}), the activity of all the mentioned radio sources suggests a possible cessation of the plasma generation in their magnetospheres \citep{GurevichIstomin2007}. This is also backed by the measured difference in spin-down rate of the intermittent pulsars PSR B1931+24 and PSR J1832+0029
in the periods of ``operation'' and ``silence'' \citep{Lyne2009}. If, alternatively, there is no plasma and no free charge escape from the stellar surface occurs, which is confirmed by the spin-down in the period of pulsar silence \citep{GurevichIstomin2007}, then the magnetosphere can be filled through the absorption of photons from the external cosmic gamma-ray background \citep{IstominSobyanin2009,IstominSobyanin2010a,IstominSobyanin2010b}. Allowance for the influence of these photons on the particle production processes was shown to be necessary in~\citep{ShukreRadhakrishnan1982}. It was pointed out in the latter paper that an influx of new charges that can be provided by the absorption of photons from the diffuse gamma-ray background is needed in the Ruderman-Sutherland model \citep{RudermanSutherland1975} to trigger each new spark in the polar gap of a pulsar magnetosphere.

Let there be a primary galactic photon whose energy and propagation direction are such that the transverse momentum component exceeds~$2m_e c$, where $m_e$ is the electron mass and $c$ is the speed of light. This photon can then produce an electron-positron pair in a magnetic field \citep{Klepikov1954}. The pair particles emit synchrotron photons as a result of the transition to the zeroth Landau level and, accelerating in a strong electric field present in the magnetosphere, begin to emit curvature photons. The absorption of these photons in a magnetic field leads to the production of the next generation of particles. The developing cascade gives rise to a lightning---a plasma tube propagating almost along magnetic field lines. In this paper, we study in detail the formation of this tube in a neutron star magnetosphere and the generation of an electron-positron plasma in it.

The paper is structured as follows. In Section~2, we provide the basic equations describing the motion of particles and the evolution of the electromagnetic field and pass to complex variables in the Maxwell equations. In Section~3, we investigate the dynamical screening of the electric field in a lightning. In Section~4, we calculate the self-consistent Lorentz factor of particles and the longitudinal electric field at the lightning center. In Sections~5 and~6, we calculate the distributions of electrons and positrons and the total number of produced electron-positron pairs, respectively, before the onset of screening and during the electric field screening. In Section~7, the calculation is concretized for a power-law dependence of the electric field on the radial coordinate. In Section~8, we estimate the radius of the forming lightning and determine the multiplicity parameter. Our main results and conclusions are summarized in Conclusions.

\section{THE MOTION OF THE PRODUCED PARTICLES AND THEIR INFLUENCE ON THE ELECTROMAGNETIC FIELD}

The motion of electrons and positrons in a neutron star magnetosphere is ultrarelativistic. Because of the presence of a strong accelerating electric field, the electrons and positrons move in opposite directions. We can write the relation
\begin{equation}
\label{electronPositronVelocity}
\mathbf{v}_\pm=\pm\mathbf{b}+\mathbf{e}\times\mathbf{b},
\end{equation}
where $\mathbf{v}_+$ is the positron velocity, $\mathbf{v}_-$ is the electron velocity, $\mathbf{b}=\mathbf{B}/B$ is a unit vector directed along a magnetic field line, and $\mathbf{e}=\mathbf{E}/B$ is the electric field vector normalized to the magnetic field strength. Formula~\eqref{electronPositronVelocity} describes the ultrarelativistic motion of charged particles along magnetic field lines and the drift motion across them. The described consideration is proper even in Deutsch's unscreened vacuum electromagnetic field \citep{Deutsch1955}, while the centrifugal drift may be neglected \citep{IstominSobyanin2009}.

The continuity equations for the electron and positron densities are
\begin{equation}
\label{continuityEquation}
\frac{\partial n_\pm}{\partial t}+\nabla\cdot n_\pm\mathbf{v}_\pm=Q(n_++n_-),
\end{equation}
where $n_+$ and $n_-$ are the positron and electron number densities, respectively, $Q$~is the total number of electron-positron pairs produced per unit time per particle. These are the basic equations for describing the generation of an electron-positron plasma.

Here, as in our previous paper \citep{IstominSobyanin2011a}, we will use a dimensionless system of units. We will measure the electric and magnetic field strengths in units of the critical field
\begin{equation*}
\label{criticalField}
B_{cr}=\frac{m_e^2c^3}{e\hbar}\approx4.414\times10^{13}\text{ G},
\end{equation*}
where $e$ is the positron charge. The length and time will be measured in units of the Compton electron wavelength $^-\!\!\!\!\lambda=\hbar/m_e c\approx3.862\times10^{-11}$~cm and its ratio to the speed of light $^-\!\!\!\!\lambda/c\approx1.288\times10^{-21}$~s, respectively. Note the useful relations $1\text{ cm}\approx2.590\times10^{10}$ and $1\text{ s}\approx7.763\times10^{20}$.

The Maxwell equations in dimensionless units in cylindrical coordinates $r,\varphi,z$ are
\begin{equation}
\label{cylindricalMaxwellEquation1}
\frac{1}{r}\frac{\partial(r E_r)}{\partial r}+\frac{\partial E_z}{\partial z}=4\pi\alpha\rho_e,
\end{equation}
\begin{equation}
\label{cylindricalMaxwellEquation2}
\frac{\partial E_\varphi}{\partial z}=\frac{\partial B_r}{\partial t},
\end{equation}
\begin{equation}
\label{cylindricalMaxwellEquation3}
\frac{\partial E_r}{\partial z}-\frac{\partial E_z}{\partial r}=-\frac{\partial B_\varphi}{\partial t},
\end{equation}
\begin{equation}
\label{cylindricalMaxwellEquation4}
\frac{1}{r}\frac{\partial(r E_\varphi)}{\partial r}=-\frac{\partial B_z}{\partial t},
\end{equation}
\begin{equation}
\label{cylindricalMaxwellEquation5}
\frac{1}{r}\frac{\partial(r B_r)}{\partial r}+\frac{\partial B_z}{\partial z}=0,
\end{equation}
\begin{equation}
\label{cylindricalMaxwellEquation6}
-\frac{\partial B_\varphi}{\partial z}=4\pi\alpha j_r+\frac{\partial E_r}{\partial t},
\end{equation}
\begin{equation}
\label{cylindricalMaxwellEquation7}
\frac{\partial B_r}{\partial z}-\frac{\partial B_z}{\partial r}=4\pi\alpha j_\varphi+\frac{\partial E_\varphi}{\partial t},
\end{equation}
\begin{equation}
\label{cylindricalMaxwellEquation8}
\frac{1}{r}\frac{\partial(r B_\varphi)}{\partial r}=4\pi\alpha j_z+\frac{\partial E_z}{\partial t}.
\end{equation}
Here, $z$ is the longitudinal coordinate measured along the magnetic field direction, $r$~is the radial coordinate, $\varphi$~is the azimuth angle, $t$~is the time, $\mathbf{E}$~and $\mathbf{B}$ are the electric and magnetic field strengths, $\rho_e=n_+-n_-$ and $\mathbf{j}=n_+\mathbf{v}_+-n_-\mathbf{v}_-$ are the charge and current densities, and $\alpha=e^2/\hbar c\approx1/137$ is the fine-structure constant. In view of the assumed local cylindrical symmetry of the plasma tube, we set $\partial/\partial\varphi\equiv0$. Since the forming plasma tube will be curved due to the existing curvature of the magnetic field lines, these equations are applicable under the condition $R\ll\rho$, where $R$ is the plasma tube radius and $\rho$ is the radius of curvature of the magnetic field lines. At the end of the paper, we show that the tube radius is limited by $100$~m (see Eq.~\eqref{radiusEstimation}), while the radius of curvature of the magnetic field lines exceeds the neutron star radius $R_S\approx10$~km. Consequently, this condition is always met.

Let us calculate the electron and positron velocities. For this purpose, let us turn to Eq.~\eqref{electronPositronVelocity}. The particle velocity is the sum of the longitudinal component giving the motion along a magnetic field line and the transverse component determined by the electric drift in crossed fields. In order of magnitude, the former and latter components are equal to $1$ and $E/B$, respectively. Initially, as long as the number of produced particles is small, the total electromagnetic field is specified by the external electric field $\mathbf{E}^{ext}$ and the external magnetic field~$\mathbf{B}^{ext}$. When, however, the plasma tube formation begins, the internal electromagnetic field also emerges. Let us specify the internal electric field $\mathbf{E}^{int}$ and the internal magnetic field~$\mathbf{B}^{int}$. All scalar components of these fields are equal between themselves in order of magnitude. The forming internal electric field will screen the external longitudinal electric field, but, obviously, it will be unable to exceed it in absolute value. We can write $B^{int}\sim E^{int}\lesssim E^{ext}$ or, passing to the components of the total field that determines the particle velocities, $B_r\sim B_\varphi\sim E_r\sim E_\varphi\sim E_z\sim E^{ext}_z$ and $B_z\sim B^{ext}_z$. In addition, we should take into account the fact that the longitudinal component $B_z$ of the total magnetic field prevails over the orthogonal components $B_r$ and~$B_\varphi$. For this reason, the total magnetic field $B$ coincides with the component $B_z$ to within terms quadratic in $B_r/B_z\sim B_\varphi/B_z$. Using these estimates, expanding the expressions in Eq.~\eqref{electronPositronVelocity} in terms of $E^{ext}_z/B^{ext}_z$ to the first order inclusive, and discarding the terms of the next orders, we will write the electron and positron velocity components
\begin{equation}
\label{particleVelocityExpansion}
v^\pm_r=\pm b_r+e_\varphi,\qquad
v^\pm_\varphi=\pm b_\varphi-e_r,\qquad
v^\pm_z=\pm 1.
\end{equation}

The particle velocities and number densities and the source of particles~$Q$, which depends on the longitudinal components of the electric and magnetic fields, enter into Eq.~\eqref{continuityEquation}. Apart from the longitudinal components, the orthogonal components of the fields, which are not required per se, enter into the Maxwell equations. To exclude them from consideration, let us introduce four new sought-for functions
\begin{equation*}
\label{Vvariables}
V^\pm_r=\pm B_r+E_\varphi,\qquad V^\pm_\varphi=\pm B_\varphi-E_r
\end{equation*}
proportional to the orthogonal particle velocity components: $V^\pm_r=B v^\pm_r$ and~$V^\pm_\varphi=B v^\pm_\varphi$. It is also convenient to pass from the variables $z,t$ to new variables
\begin{equation*}
\label{tPlusMinusVariables}
t_\pm= t\pm z.
\end{equation*}
We can now rewrite the Maxwell equations \eqref{cylindricalMaxwellEquation1}--\eqref{cylindricalMaxwellEquation8} by specifying the functions $V^+_r$, $V^-_r$, $V^+_\varphi$, $V^-_\varphi$, $E_z$, and $B_z$ dependent on the variables $t_+,t_-,r$ as unknown quantities:
\begin{equation}
\label{cylindricalMaxwellEquation1onV}
2\frac{\partial V^\pm_r}{\partial t_\mp}=-4\pi
\alpha j_\varphi-\frac{\partial B_z}{\partial r},
\end{equation}
\begin{equation}
\label{cylindricalMaxwellEquation2onV}
\frac{1}{r}\frac{\partial(r V^\pm_r)}{\partial r}=
-2\frac{\partial B_z}{\partial t_\pm},
\end{equation}
\begin{equation}
\label{cylindricalMaxwellEquation3onV}
2\frac{\partial V^\pm_\varphi}{\partial t_\mp}=
4\pi\alpha j_r \pm\frac{\partial E_z}{\partial r},
\end{equation}
\begin{equation}
\label{cylindricalMaxwellEquation4onV}
\frac{1}{r}\frac{\partial(r V^\pm_\varphi)}{\partial r}=
\pm 4\pi\alpha(j_z\mp\rho_e) \pm2\frac{\partial E_z}{\partial t_\pm}.
\end{equation}

Let us introduce a complex electromagnetic field vector $\mathbf{F}=\mathbf{E}+i\mathbf{B}$, a complex transverse velocity $v^\pm_\perp= v^\pm_r+iv^\pm_\varphi$, and a complex transverse electric current density $j_\perp= j_r+ij_\varphi$. Equations~\eqref{cylindricalMaxwellEquation1onV}--\eqref{cylindricalMaxwellEquation4onV} will then take a more compact form:
\begin{equation*}
\label{complexCylindricalMaxwellEquation1}
2\frac{\partial V^\pm_\perp}{\partial t_\mp}=
i\left(4\pi\alpha j_\perp \pm\frac{\partial F^\pm_z }{\partial r}\right),
\end{equation*}
\begin{equation}
\label{complexCylindricalMaxwellEquation2}
\frac{1}{r}\frac{\partial(rV^\pm_\perp)}{\partial r}=
\pm 2i\left(4\pi\alpha n_\mp +\frac{\partial F^\pm_z }{\partial t_\pm}\right),
\end{equation}
where we introduced the quantity $V^\pm_\perp= Bv^\pm_\perp$ and the longitudinal component $F^+_z\equiv F_z$ of the complex electromagnetic field vector and the corresponding complex conjugate component $F^-_z= F^*_z$. It is also easy to derive the equations for the longitudinal components $F^\pm_z$ that we will need below:
\begin{equation}
\label{dalambertianF}
\Box\,F^\pm_z=4\pi\alpha\left(4\frac{\partial n_\mp}
{\partial t_\mp} \mp\frac{1}{r}\frac{\partial(rj_\perp)}{\partial r}\right),\qquad
\Box=\triangle-\partial^2/\partial t^2.
\end{equation}
They are convenient in that they contain the dependence of the longitudinal component of the complex electromagnetic field only on the electron and positron densities and the complex transverse current density. However, the transverse current is not specified independently but is expressed in terms of the particle densities and velocities as follows:
\begin{equation}
\label{complexOrthogonalCurrent}
j_\perp=\frac{1}{B}\bigl(n_+V^+_\perp-n_-V^-_\perp\bigr).
\end{equation}

Below, we will expand all of the quantities entering into the equations in terms of the radial coordinate~$r$. If we arbitrarily designate some sought-for quantity as~$\mathcal{Y}$, then its expansion in general form is represented by a power series
\begin{equation*}
\label{Yseries}
\mathcal{Y}=\sum\limits_{n=0}^\infty \mathcal{Y}^{(n)}\frac{r^n}{n!},
\end{equation*}
where the quantities
\begin{equation*}
\label{Yderivatives}
\mathcal{Y}^{(n)}=\left.\frac{\partial^n \mathcal{Y}}{\partial r^n}\right|_{r=0}
\end{equation*}
depend only on the coordinates $z,t$ or $t_+,t_-$ and do not depend on~$r$. Obviously, the positron and electron densities, the charge and current densities, and the longitudinal electromagnetic field components are expanded in terms of even powers of~$r$,  while the transverse components of the electromagnetic field and the particle velocities are expanded in terms of odd powers of~$r$. For example,
\begin{equation*}
\label{radialExpansion}
F_z=F^{(0)}_z+F^{(2)}_z \frac{r^2}{2}+O(r^4),\qquad
n_\pm=n^{(0)}_\pm+O(r^2),\qquad
v^\pm_\perp=v^{\pm(1)}_\perp r+O(r^3).
\end{equation*}
Expanding all of the quantities entering into the equation into series of $r$ and equating the terms with the same powers of~$r$, we will obtain an infinite set of equations among which we will restrict our consideration to the lowest-order equation.

To find the expansion of the current density~$j_\perp$, we will need the expansion of Eq.~\eqref{complexCylindricalMaxwellEquation2}:
\begin{equation*}
\label{expandedComplexCylindricalMaxwellEquation2}
V^{\pm(1)}_\perp=\pm i\left(4\pi\alpha n^{(0)}_\mp +\frac{\partial F^{\pm(0)}_z }{\partial t_\pm}\right).
\end{equation*}
Now, using Eq.~\eqref{complexOrthogonalCurrent}, we easily obtain
\begin{equation*}
\label{expandedComplexCurrent}
j^{(1)}_\perp=\frac{i}{B^{(0)}}\left(n^{(0)}_+\frac{\partial F^{+(0)}_z}{\partial t_+} +n^{(0)}_-\frac{\partial F^{-(0)}_z}{\partial t_-}+8\pi\alpha n^{(0)}_+n^{(0)}_-\right).
\end{equation*}

The continuity equations for finding the positron and electron densities
\begin{equation}
\label{continuityEquationsInNewVariables}
2\frac{\partial n_\pm}{\partial t_\pm}+\frac{1}{r}\frac{\partial(rn_\pm v^\pm_r)}{\partial r}=Qj_z
\end{equation}
after the expansion take the form
\begin{equation}
\label{expandedContinuityEquation}
\frac{\partial n^{(0)}_\pm}{\partial t_\pm}+n^{(0)}_\pm v^{\pm(1)}_r
=\frac{1}{2}Q^{(0)}j^{(0)}_z.
\end{equation}
The equation for the longitudinal electromagnetic field component follows from the expansion of Eq.~\eqref{dalambertianF}:
\begin{equation}
\label{expandedDalabbertianF}
2F^{(2)}_z+\Box_\parallel\,F^{(0)}_z=8\pi\alpha\left(\frac{\partial n^{(0)}_+}{\partial t_+}+\frac{\partial n^{(0)}_-}{\partial t_-}-ij^{(1)}_\varphi\right),
\end{equation}
where we introduced the D'Alembert operator in longitudinal coordinate
\begin{equation*}
\label{longitudinalDalambertian}
\Box_\parallel=\frac{\partial^2}{\partial z^2}-\frac{\partial^2}{\partial t^2}.
\end{equation*}
Now, it remains to express the derivatives of the positron and electron densities in terms of the particle source and the radial fluxes using Eqs.~\eqref{expandedContinuityEquation} and to substitute the derived expressions into Eq.~\eqref{expandedDalabbertianF}. Separating out the imaginary part of the complex current density $j_\perp$ allows the azimuthal current density to be obtained:
\begin{equation*}
\label{expandedJphi}
j^{(1)}_\varphi=\frac{1}{B^{(0)}}\left(n^{(0)}_+\frac{\partial E^{(0)}_z}{\partial t_+} +n^{(0)}_-\frac{\partial E^{(0)}_z}{\partial t_-}+8\pi\alpha n^{(0)}_+n^{(0)}_-\right).
\end{equation*}
The expansions of the radial positron and electron velocities can be derived from the relation $v^{\pm(1)}_r=V^{\pm(1)}_r/B^{(0)}$, once the real part of $V^{\pm(1)}_\perp$ has been separated out:
\begin{equation*}
\label{expandedVr}
V^{\pm(1)}_r=-\frac{\partial B^{(0)}_z}{\partial t_\pm}.
\end{equation*}
Finally, we obtain the equation for the longitudinal component of the complex electromagnetic field vector
\begin{equation}
\label{FviaQandNumberDensities}
2F^{(2)}_z+\Box_\parallel\,F^{(0)}_z=8\pi\alpha\biggl[Q^{(0)}j^{(0)}_z -\frac{i}{B^{(0)}}\left(n^{(0)}_+\frac{\partial F^{(0)}_z}{\partial t_+} +n^{(0)}_-\frac{\partial F^{(0)}_z}{\partial t_-}+8\pi\alpha n^{(0)}_+n^{(0)}_-\right)\biggr].
\end{equation}
This equation is convenient in that it contains only the longitudinal components $E^{(0)}_z$ and $B^{(0)}_z$ of the electromagnetic field and the electron and positron densities. The radial and azimuthal components of the electromagnetic field and the particle velocities turned out to be excluded.

\section{DYNAMICAL SCREENING OF THE LONGITUDINAL ELECTRIC FIELD}

It was shown in \citep{SobyaninDissertation} that the first term on the right-hand side of Eq.~\eqref{FviaQandNumberDensities} prevails over the remaining terms. Note that this can be verified only after the calculation of the self-consistent source of electron-positron pairs by taking into account the dynamical screening of the external longitudinal electric field. Let us turn to the left-hand side of Eq.~\eqref{FviaQandNumberDensities}. Initially, this expression is equal to~$\Box\,F_z$. The D'Alembert operator is represented as the sum
\begin{equation*}
\label{DalambertianAsAsum}
\Box=\triangle_\perp+\Box_\parallel,
\end{equation*}
where the two-dimensional Laplace operator in radial coordinate was introduced:
\begin{equation*}
\label{LaplacianPerp}
\triangle_\perp=\frac{1}{r}\frac{\partial}{\partial r} r\frac{\partial}{\partial r}.
\end{equation*}
Let the characteristic longitudinal distance and the characteristic time scale on which the electromagnetic field changes be $Z$ and $T$, respectively. The field changes with radial coordinate on distances of the order of the radius $R$ of the forming plasma tube. We can then symbolically write
\begin{equation*}
\label{LaplacianPerpAndDalambertianParallel}
\triangle_\perp\sim\frac{1}{R^2},\qquad
\Box_\parallel\sim\frac{1}{Z^2}-\frac{1}{T^2}.
\end{equation*}
We will assume that, first, the change of the electromagnetic field is quasi-stationary and, second, the characteristic longitudinal scales of the field change exceed considerably the corresponding transverse scales. Then, this means that $\Box_\parallel\ll\triangle_\perp$. The stipulated conditions allow us to formally set $\Box_\parallel=0$ in Eq.~\eqref{FviaQandNumberDensities} and to obtain the final equation for the longitudinal component of the complex electromagnetic field:
\begin{equation*}
\label{finalEquationForComplexFieldComponent}
F^{(2)}_z=4\pi\alpha Q^{(0)}j^{(0)}_z.
\end{equation*}
Once the real and imaginary parts have been separated out, we have
\begin{equation}
\label{finalEqEandB}
E^{(2)}_z=4\pi\alpha Q^{(0)}j^{(0)}_z,\qquad
B^{(2)}_z=0.
\end{equation}

The radial distribution of the longitudinal electric field inside the plasma tube is
\begin{equation*}
\label{parallelElectricFieldRadialExpansion}
E_z\approx E^{(0)}_z+E^{(2)}_z\frac{r^2}{2},\qquad r\leqslant R.
\end{equation*}
Let us specify some electric field $E^{bound}_z= E_z|_{r=R}$ at the tube boundary. To determine this field, we should take into account the fact that the longitudinal electric field at infinity is equal to the external one, $E_z|_{r=\infty}=E^{ext}_z$. Next, let us turn to Eq.~\eqref{cylindricalMaxwellEquation3}. The radial derivative $\partial E_z/\partial r$ entering into it exceeds the derivatives $\partial E_r/\partial z$ and $\partial B_\varphi/\partial t$ in order of magnitude. This follows from the field estimates pointed out after Eq.~\eqref{cylindricalMaxwellEquation8} and from
the inequalities $Z\gg R$ and~$R\ll T$. We arrive at the equality $\partial E_z/\partial r=0$ whose integration leads us to the boundary condition $E^{bound}_z=E^{ext}_z$. Thus, the longitudinal electric field at the plasma tube boundary is approximately equal to the external longitudinal electric field existing in the magnetosphere in the absence of an electron-positron plasma. The longitudinal electric field at the tube center is then
\begin{equation}
\label{EcenterViaEext}
E^{(0)}_z=E^{ext}_z-E^{(2)}_z\frac{R^2}{2}.
\end{equation}
Similar reasoning is also valid for the longitudinal magnetic field. We should only turn to Eq.~\eqref{cylindricalMaxwellEquation7} and take into account the fact that $j_\varphi=0$ outside the plasma tube. The boundary magnetic field $B^{bound}_z= B_z|_{r=R}$ then coincides with the external longitudinal magnetic field~$B^{ext}_z$. The equality $B^{(0)}_z=B^{bound}_z$ or, alternatively,
\begin{equation}
\label{BcenterViaBext}
B^{(0)}_z=B^{ext}_z
\end{equation}
follows from Eq.~\eqref{finalEqEandB}.

Consider a quasi-stationary plasma tube. Obviously, the longitudinal electric current existing in the plasma tube gives rise to a nonzero azimuthal magnetic field~$B_\varphi$. Since the particles move mainly along magnetic field lines, the appearance of an azimuthal magnetic field component leads to the appearance of an azimuthal particle velocity component~$v_\varphi$. Here, in general, the existence of an additional drift velocity component due to the presence of a radial electric field $E_r$ should also be taken into account (see Eq.~\eqref{particleVelocityExpansion}). This is unimportant for qualitative considerations, because we can consider the regions in the plasma tube where the electron and positron densities are comparable between themselves but fairly high per se. The radial electric field will then be low, while a certain electric current causing the magnetic field lines to twist will flow in the tube. An azimuthal electric current component $j_\varphi$ will appear, which, in turn, will cause the longitudinal magnetic field to change. This change will be nonuniform in radial coordinate. Consequently, the second derivative of the magnetic field is still nonzero. However, it is small compared to the second derivative of the longitudinal electric field. This is because the production of a secondary electron-positron plasma is fairly efficient due to the large~$Q$. Therefore, Eq.~\eqref{BcenterViaBext} should be understood in the sense that the change in longitudinal magnetic field through the appearance of an internal screening electromagnetic field in the plasma tube is small compared to the change in longitudinal electric field caused by the same internal field.

Now, returning to Eq.~\eqref{finalEqEandB}, we can understand the longitudinal electric field screening mechanism. Here, it is important that the electron-positron pair production rate primarily determines the second spatial derivatives of the longitudinal electric field. In other words, the curvature of the two-dimensional cylindrically symmetric surface $E_z(r)$ increases with plasma generation rate. Since the boundary condition $E^{bound}_z=E^{ext}_z$ is met, the surface sag causes the longitudinal electric field at the plasma tube center to decrease. This mechanism may be called dynamical screening, because the surface sag depends on the plasma generation rate. It differs from the mechanism of ordinary static screening related to charge separation. The existence of screening different from static one was shown in~\citep{BeskinScr1982}. The electric field screening in the sparks forming in the polar gap considered in the latter paper was also called dynamical one.

Let us integrate Eq.~\eqref{continuityEquationsInNewVariables} over the radial coordinate with the weight $\pi r$ from~$0$ to~$\infty$:
\begin{equation}
\label{integratedContinuityEq}
\frac{\partial N_\pm}{\partial t_\pm}=\pi\int\limits_0^R{Qj_zrdr}.
\end{equation}
Here, $N_+$ and $N_-$ denote the linear positron and electron densities, respectively. As above, let us expand all quantities in terms of~$r$. Let us also introduce the linear charge density $N_\rho= N_+-N_-$ and the electric current $J_z= N_++N_-$. We can then write the relations
\begin{equation}
\label{linearDensities}
N_\pm=n^{(0)}_\pm\pi R^2,\qquad
N_\rho=\rho_e^{(0)}\pi R^2,\qquad
J_z=j^{(0)}_z\pi R^2.
\end{equation}
Equation \eqref{integratedContinuityEq} transforms to
\begin{equation}
\label{expandedIntegratedContinuityEq}
\frac{\partial N_\pm}{\partial t_\pm}=\frac{Q^{(0)}J_z}{2}.
\end{equation}
We can also pass from the variables $N_\pm$ to the variables $N_\rho$ and $J_z$ and obtain
\begin{equation}
\label{expandedEqForRho}
\frac{\partial N_\rho}{\partial t}+\frac{\partial J_z}{\partial z}=0,
\end{equation}
\begin{equation}
\label{expandedEqForJz}
\frac{\partial J_z}{\partial t}+\frac{\partial N_\rho}{\partial z} =2Q^{(0)}J_z.
\end{equation}

Thus, to find the distribution of electrons and positrons along the longitudinal coordinate and to calculate the number of particles produced in the plasma tube lifetime, the source of electron-positron pairs per particle $Q^{(0)}$ should be found.

\section{THE PARTICLE LORENTZ FACTOR}

Formulas \eqref{finalEqEandB}, \eqref{EcenterViaEext}, and \eqref{linearDensities} allow the dynamical screening of the longitudinal electric field to be taken into account:
\begin{equation*}
\label{electricFieldWithScreening}
E^{(0)}_z=E^{ext}_z-2\alpha Q^{(0)}J_z.
\end{equation*}
As $Q^{(0)}$, we should take the effective local source
\begin{equation}
\label{finalQeff}
Q^{eff}=\frac{\alpha}{3\,\xi_0}\frac{\gamma_0^3}{\rho^2},
\end{equation}
calculated for the plasma tube center \citep{IstominSobyanin2011a}. The source $Q^{eff}$ depends on the logarithmic factor $\xi_0\sim0.1$ and the stationary Lorentz factor
\begin{equation}
\label{gammaMax}
\gamma_0=\left(\frac{3}{2\alpha}E_\parallel\rho^2\right)^{1/4},
\end{equation}
which, in turn, is a function of the longitudinal electric field $E_\parallel\equiv E_z$. For the tube center, we have $E_z=E^{(0)}_z$. The described dependence allows the Lorentz factor $\gamma_0$ and the longitudinal electric field $E^{(0)}_z$ at the plasma
tube center to be calculated self-consistently. The Lorentz factor $\gamma_0$ can be found from the equation
\begin{equation}
\label{forthDegreeEq}
\gamma_0^4+\frac{\alpha}{\xi_0}J_z\gamma_0^3-\gamma_{ext}^4=0,
\end{equation}
where $\gamma_{ext}$ is defined by Eq.~\eqref{gammaMax}, in which we should set $E_\parallel=E_z^{ext}$.

Equation~\eqref{forthDegreeEq} is a quartic algebraic equation. It can be brought to the form (for more detail, see \citep{Faddeev1984})
\begin{equation}
\label{factorizedForthDegreeEq}
\begin{split}
\biggl(\gamma_0^2&+\Bigl(\frac{\alpha}{2\xi_0}J_z-K\Bigr)\gamma_0+\Bigl(\frac{1}{2}\,Y-L\Bigr)\!\biggr)
\\
\times
\biggl(\gamma_0^2&+\Bigl(\frac{\alpha}{2\xi_0}J_z+K\Bigr)\gamma_0+\Bigl(\frac{1}{2}\,Y+L\Bigr)\!\biggr)=0,
\end{split}
\end{equation}
where we can take
\begin{equation*}
\label{exprK}
K=\sqrt{\Bigl(\frac{\alpha}{2\xi_0}J_z\Bigr)^2+Y},\qquad
L=-\sqrt{\frac{1}{4}\,Y^2+\gamma_{ext}^4}
\end{equation*}
as $K$ and $L$. Here, $Y$ is one of the solutions of the cubic equation
\begin{equation}
\label{thirdDegreeEq}
Y^3+D_1Y+D_2=0,
\end{equation}
where
\begin{equation*}
\label{D1andD2}
D_1=4\gamma_{ext}^4,\qquad
D_2=\Bigl(\frac{\alpha}{2\xi_0}J_z\Bigr)^2D_1.
\end{equation*}
Since $D_1>0$, this equation has one real and two complex conjugate roots. Let $Y$ mean the real root of Eq.~\eqref{thirdDegreeEq}. Cardano's formula gives
\begin{equation*}
\label{Cardano}
Y=\sqrt[3]{-\frac{D_2}{2}+\sqrt{\frac{D_2^2}{4}+\frac{D_1^3}{27}}}
-\sqrt[3]{\frac{D_2}{2}+\sqrt{\frac{D_2^2}{4}+\frac{D_1^3}{27}}},
\end{equation*}
where the real values should be taken for the left and right cubic roots. Obviously, $Y<0$.

Let us return to Eq.~\eqref{factorizedForthDegreeEq}. The quadratic equation obtained when the left bracket is equal to zero will not be of interest to us, because, as is easy to verify, it has a negative discriminant. This corresponds to the existence of two complex conjugate roots. The quadratic equation obtained when the right bracket is set equal to zero has two real roots of opposite signs. Choosing the positive root, we obtain the required expression for the Lorentz factor:
\begin{equation}
\label{finalGamma0}
\gamma_0=\frac{1}{2}\Bigl(\frac{\alpha}{2\xi_0}J_z+K\Bigr)
\biggl(-1+\sqrt{1-\Bigl(\frac{Y}{K}\Bigr)^2\Bigl(\frac{1}{2}\,Y+L\Bigr)^{-1}}\biggr).
\end{equation}
The corresponding total longitudinal electric field at the plasma tube center is
\begin{equation}
\label{centerElectricField}
E^{(0)}_z=\frac{2}{3}\alpha\frac{\gamma_0^4}{\rho^2}.
\end{equation}

Since Eq.~\eqref{finalGamma0} is difficult to use, let us construct its acceptable approximation. For this purpose, consider the behavior of the Lorentz factor $\gamma_0$ at low and high currents~$J_z$. If the current $J_z$ is low, then we can neglect the second term in Eq.~\eqref{forthDegreeEq} and obtain
\begin{equation}
\label{gamma0AtLowCurrent}
\gamma_0=\gamma_{ext},\qquad J_z\rightarrow0.
\end{equation}
As it must be, the screening of the external longitudinal electric field disappears at low currents. If, alternatively, the current $J_z$ is high, then we can neglect the first term in Eq.~\eqref{forthDegreeEq}:
\begin{equation}
\label{gamma0AtHighCurrent}
\gamma_0=\gamma_{ext}\Bigl(\frac{J_z}{J_0}\Bigr)^{-1/3},\qquad J_z\gg J_0.
\end{equation}
Here, we introduced the screening current
\begin{equation}
\label{screeningCurrent}
J_0=\frac{\xi_0}{\alpha}\gamma_{ext},
\end{equation}
in exceeding which the screening of the longitudinal electric field becomes significant. If $\gamma_{ext}\sim10^8$ and $\xi_0\sim0.1$, then $J_0\sim10^9$. To construct the approximation, we will proceed from Eq.~\eqref{gamma0AtHighCurrent}. Its shortcoming lies in the existence of a divergence when $J_z\rightarrow0$. To satisfy requirement~\eqref{gamma0AtLowCurrent}, let us shift the argument $J_z$ in Eq.~\eqref{gamma0AtHighCurrent} by the screening current~$J_0$:
\begin{equation}
\label{gamma0approximation}
\gamma^{ap}_0\approx\gamma_{ext}\biggl(1+\frac{J_z}{J_0}\biggr)^{-1/3}.
\end{equation}
It is easy to see that this approximation satisfies asymptotics \eqref{gamma0AtLowCurrent} and \eqref{gamma0AtHighCurrent}. The accuracy of the approximation can be estimated by calculating $(\gamma_0-\gamma^{ap}_0)/\gamma_0$, where $\gamma_0$ is the exact solution~\eqref{finalGamma0}. Obviously, this quantity depends only on the ratio~$J_z/J_0$. The dependence is plotted in the figure. We see that the approximation is rather accurate and the maximum deviation of $\gamma^{ap}_0$ from $\gamma_0$ reaches~$3.3\%$ at~$J_z/J_0\approx1.62$. This allows the approximation $\gamma_0^{ap}$ to be used instead of the exact solution $\gamma_0$ in the subsequent calculations.

Using Eq.~\eqref{centerElectricField}, we can derive the dependence of the longitudinal electric field $E^{(0)}_z$ on the longitudinal current~$J_z$:
\begin{equation*}
\label{centerEzWithScreening}
E^{(0)}_z=E^{ext}_z\biggl(1+\frac{J_z}{J_0}\biggr)^{-4/3}.
\end{equation*}
This dependence shows that the screening is switched on at currents $J_z\sim J_0$. Now, using Eq.~\eqref{finalQeff}, we obtain
\begin{equation}
\label{centerQwithScreening}
Q^{(0)}=\frac{Q_0}{1+J_z/J_0},
\end{equation}
where we introduced the source
\begin{equation}
\label{Qext}
Q_0= \left.Q^{eff}\right|_{\gamma_0=\gamma_{ext}},
\end{equation}
equal to the effective source of particles when there is no screening of the external longitudinal electric field. Let~$E_\parallel\sim10^{-6}$ and~$\gamma_{ext}\sim 10^8$. The particle acceleration time to the stationary Lorentz factor is then $\tau_{st}=\gamma_{ext}/E_\parallel\sim10^{14}$, which gives an estimate of $Q_0\sim10^{-13}$ at~$\xi_0\sim0.1$.

\section{PARTICLE PRODUCTION BEFORE SCREENING}

Let us find the longitudinal distribution of the electron and positron densities and the corresponding total number of produced particles. For this purpose, we will take into account dependence~\eqref{centerQwithScreening} and rewrite Eq.~\eqref{expandedEqForJz} as
\begin{equation}
\label{eqForJzWithQ0}
\frac{\partial J_z}{\partial t}+\frac{\partial N_\rho}{\partial z}
=\frac{2Q_0J_z}{1+J_z/J_0}.
\end{equation}
We will not write Eq.~\eqref{expandedEqForRho}, which is an obvious corollary of the charge conservation law. The total source of particles, i.e., the total number of electrons and positrons produced per unit time in a plasma tube segment with a unit length, appears on the right-hand side of Eq.~\eqref{eqForJzWithQ0}. At a low longitudinal current~$J_z$, Eq.~\eqref{eqForJzWithQ0} transforms to
\begin{equation*}
\label{expandedEqForSmallJz}
\frac{\partial J_z}{\partial t}+\frac{\partial N_\rho}{\partial z}=2Q_0J_z,
\qquad J_z\ll J_0,
\end{equation*}
i.e., the electron-positron pair production rate is the same as that in an unscreened external electromagnetic field. If, alternatively, the current exceeds~$J_0$, then
\begin{equation}
\label{expandedEqForBigJz}
\frac{\partial J_z}{\partial t}+\frac{\partial N_\rho}{\partial z}=2Q_0J_0,
\qquad J_z\gg J_0.
\end{equation}
In this case, the pair production rate reaches some saturation value and the dependence on the total longitudinal current disappears. Hence, at a high longitudinal current $J_z\gg J_0$, when the longitudinal electric field screening processes come to the fore, the total electron-positron pair production rate coincides with that in the absence of screening but with a fixed longitudinal current $J_z=J_0$. Thus, here we observe the saturation of the particle source and the critical current $J_0$ may be called the saturation one.

By differentiating Eq.~\eqref{eqForJzWithQ0} with respect to time by taking into account the charge conservation law~\eqref{expandedEqForRho}, we can exclude the linear charge density from consideration and consider the equation for the longitudinal current~$J_z$:
\begin{equation*}
\label{linearLongitudinalCurrentEq}
\Box_\parallel J_z=-\frac{2Q_0}{(1+J_z/J_0)^2}\frac{\partial J_z}{\partial t}.
\end{equation*}
For this equation, we should consider the Cauchy problem with the initial conditions
\begin{equation*}
\label{initialConditions}
\left.J_z\right|_{t=0}=J^{init}_z,\qquad\left.\frac{\partial J_z}{\partial t}\right|_{t=0}=
\frac{2Q_0J^{init}_z}{1+J^{init}_z/J_0}-\frac{\partial N^{init}_\rho}{\partial z},
\end{equation*}
where
\begin{equation*}
\label{RhoInit}
N^{init}_\rho=\left.N_\rho\right|_{t=0}.
\end{equation*}
The initial linear charge density $N^{init}_\rho$ and the initial current $J^{init}_z$ are some arbitrary functions of the longitudinal coordinate~$z$.

Let initially $J^{init}_z\ll J_0$. As long as $J_z<J_0$, the plasma multiplication will be described by a linear hyperbolic equation:
\begin{equation*}
\label{hyperbolicEqSmallJz}
\Box_\parallel J_z=-2Q_0\frac{\partial J_z}{\partial t},\qquad
J_z\ll J_0.
\end{equation*}
Its solution is
\begin{equation}
\label{smallJzSolution}
\begin{split}
J_z(z,t)=\frac{e^{Q_0t}}{2}\Biggl[\:\int\limits_{z-t}^{z+t}I_0\bigl(Q_0\sqrt{t^2-(z-z')^2}\bigr)
\left(\left.\frac{\partial J_z}{\partial t}\right|_{t=0}(z')-Q_0\!\left.J_z\right|_{t=0}(z')\right)dz'\\
+\int\limits_{z-t}^{z+t}\frac{Q_0t}{\sqrt{t^2-(z-z')^2}}\,I_1\bigl(Q_0\sqrt{t^2-(z-z')^2}\bigr)
\left.J_z\right|_{t=0}(z')\,dz'\\
+\left.J_z\right|_{t=0}(z-t)+\left.J_z\right|_{t=0}(z+t)\Biggr],
\end{split}
\end{equation}
where $I_0(z)$ and $I_1(z)$ are, respectively, the zeroth- and first-order Bessel functions of an imaginary argument. Using solution \eqref{smallJzSolution} and the charge conservation law~\eqref{expandedEqForRho}, we obtain the linear charge density
\begin{equation}
\label{smallJzSolutionForRho}
N_\rho(z,t)=N^{init}_\rho(z)-\int\limits_0^t\frac{\partial J_z(z,t')}{\partial z}\,dt'.
\end{equation}
Hence we can derive the expressions for the linear electron and positron densities
\begin{equation}
\label{NplusMinusViaJzAndRho}
N_\pm=\frac{J_z\pm N_\rho}{2}.
\end{equation}
For the case under consideration, we should take
\[
J^{init}_z=N^{init}_++N^{init}_-,\qquad N^{init}_\rho=N^{init}_+-N^{init}_-,
\]
where $N^{init}_\pm$ is defined by the formula
\begin{equation}
\label{electronAndPositronDensityInitialConditions}
N^{init}_\pm=\frac{3}{2l_1}\theta(l_1-|z|).
\end{equation}
Here, $\theta(x)$ is the theta function and $l_1$ is the distance from the primary electron-positron pair produced by a photon from the external cosmic gamma-ray background at which secondary electron-positron pairs are produced \citep{IstominSobyanin2011a}.

It makes sense to consider solution \eqref{smallJzSolution} at currents $J_z$ that are low compared to the screening current~$J_0$. To estimate the time until which this can be done, note that, being symmetric relative to the initial point~$z=0$, solution~\eqref{smallJzSolution} at long times has a maximum at this point. The required time that we will denote by~$t_{scr}$ can then be determined from the equation
\begin{equation}
\label{exprForTscr}
J_z(0,t_{scr})=J_0.
\end{equation}
We will assume that $t\gg \tau_1\approx l_1$, where $\tau_1$ is the time corresponding to the distance~$l_1$. The equality
\begin{equation}
\label{smallJzSolutionWhenTggL1}
J_z(0,t)=3Q_0 e^{Q_0t}\Bigl(I_0(Q_0t)+I_1(Q_0t)\Bigr),
\qquad t\gg \tau_1,
\end{equation}
then follows from Eq.~\eqref{smallJzSolution}. We will also assume that $Q_0t\gg1$. Under the condition $\tau_1\sim\tau_{st}$, it follows from Eq.~\eqref{finalQeff} that $t\gg\tau_1\xi_0$. Since~$\xi_0\ll1$, this condition does not restrict us in any way if $t\gg \tau_1$. The asymptotics of the Bessel functions of an imaginary argument \citep{NikiforovUvarov}
\begin{equation*}
\label{imaginaryBesselAsymptotics}
I_\nu(z)=\frac{e^z}{\sqrt{2\pi z}},\qquad z\rightarrow\infty,
\end{equation*}
then allows Eq.~\eqref{smallJzSolutionWhenTggL1} to be rewritten as
\begin{equation*}
\label{smallJzSolutionWhenQ0Tgg1}
J_z(0,t)=6\sqrt{\frac{Q_0}{2\pi t}}\,e^{2Q_0 t},\qquad Q_0 t\gg1.
\end{equation*}
Now, Eq.~\eqref{exprForTscr} can be brought to the equivalent form
\begin{equation}
\label{equivEqForTscr}
\xi_{scr}=\Lambda_{scr}+\frac{1}{2}\ln\xi_{scr},
\end{equation}
where
\begin{equation*}
\label{LambdaScr}
\Lambda_{scr}=\ln\frac{J_0\sqrt{\pi}}{6Q_0}.
\end{equation*}
At $J_0\sim10^9$ and $Q_0\sim10^{-13}$, we have $\Lambda_{scr}\approx50$. At~$\Lambda_{scr}\gg1$, Eq.~\eqref{equivEqForTscr} has an approximate solution,
\begin{equation*}
\label{approxXiScr}
\xi_{scr}\approx\Lambda_{scr}+\frac{1}{2}\ln\Lambda_{scr}.
\end{equation*}
For the source $Q_0$ and the linear longitudinal screening current density $J_0$ mentioned above, we have $\xi_{scr}\approx52$, which corresponds to $Q_0t_{scr}\gg1$. The sought-for screening time is then
\begin{equation}
\label{screeningTime}
t_{scr}=\frac{\xi_{scr}}{2Q_0},
\end{equation}
with $t_{scr}\sim10^{14}-10^{15}$. The ratio
\begin{equation*}
\label{tScrToTauSt}
\frac{t_{scr}}{\tau_{st}}=\xi_0\xi_{scr}
\end{equation*}
is approximately equal to $t_{scr}/\tau_{st}\approx5>1$. It remains to determine the total number $N_{\Sigma}^{scr}$ of electron-positron pairs produced before screening. Integrating the equation
\begin{equation*}
\label{finaldNdt}
\frac{dN_\Sigma(t)}{dt}=2Q_0N_\Sigma(t),
\end{equation*}
written for the total number of electron-positron pairs,
\begin{equation}
\label{notationForIntegratedDensities}
N_\Sigma(t)=\int\limits_{-\infty}^\infty N_\pm(z,t)dz,
\end{equation}
and choosing $N_{\Sigma}^{init}=3$ as the initial condition (see Eqs.~\eqref{electronAndPositronDensityInitialConditions} and~\eqref{notationForIntegratedDensities}), we obtain
\begin{equation*}
\label{totalParticleNumberBeforeScreening}
N_{\Sigma}^{scr}=3e^{2Q_0 t_{scr}}.
\end{equation*}
Finally, using Eqs.~\eqref{equivEqForTscr} and~\eqref{screeningTime}, we can derive the final expression
\begin{equation}
\label{finalTotalParticleNumberBeforeScreening}
N_{\Sigma}^{scr}=\sqrt{\pi\xi_{scr}}\frac{J_0}{2Q_0}.
\end{equation}
At $J_0\sim10^9$, $Q_0\sim10^{-13}$, and $\xi_0\sim0.1$, we obtain $N_{\Sigma}^{scr}\sim10^{22}$.

\section{PARTICLE PRODUCTION DURING SCREENING}

After the time $\tau_{scr}$, the total source of electron-positron pairs becomes constant due to the saturation effect (see Eq.~\eqref{expandedEqForBigJz}). However, we should take into account the fact that this source is constant only on distance scales that are small compared to the scales of the change in external magnetospheric electromagnetic field coincident with the neutron star radius. In the preceding section, this did not require a special allowance, because $t_{scr}\ll R_S$. In contrast, the plasma generation processes will subsequently develop on scales of the order of $R_S$---the scales that the plasma tube length will eventually reach. Let us introduce the time
\begin{equation*}
\label{t0time}
t_0=\tau_1+t_{scr},
\end{equation*}
after which (from the appearance of a primary electron-positron pair) the plasma generation will occur under saturation. It can be assumed that $t_0\approx t_{scr}$. Let us designate the total source of electron-positron pairs as
\begin{equation*}
\label{QsumAsQ0J0}
Q_\Sigma(z)= Q_0J_0.
\end{equation*}
We will assume it to be some arbitrary function of the longitudinal coordinate~$z$. This can be done, because $Q_\Sigma(z)$ depends only on the external parameters of the
magnetospheric electromagnetic field but not on the electron and positron densities in the plasma tube. Recall that the production point of a primary electron-positron pair fixes some magnetic field line. The produced secondary particles will move almost along the field line under consideration and their positions can be specified by some longitudinal coordinate~$z$. Fixing the field line for known external electromagnetic fields completely determines the function~$Q_\Sigma(z)$. Let us also introduce the longitudinal coordinate $z_0$ corresponding to the production point of a primary electron-positron pair. Without loss of generality, we will set $z_0=0$. To calculate the linear electron and positron densities, it is more convenient to directly use Eq.~\eqref{expandedIntegratedContinuityEq} and to write
\begin{equation}
\label{pairGenerationWhenScreening}
\frac{\partial N_\pm}{\partial t_\pm}=\frac{1}{2}Q_\Sigma(z)\theta(t_0+t-|z|).
\end{equation}
In this case, we consider $t_0$ to be the initial time to be able to write the initial conditions derived from the calculations of the preceding section as the conditions at $t=0$. Direct integration of Eqs.~\eqref{pairGenerationWhenScreening} gives the linear densities of positrons
\begin{equation*}
\label{positronLinearDensityWithScreening}
\begin{split}
N_+(z,t)=\left.N_+\right|_{t=0}(-t_-)&+\biggl[\theta(t_0+t_+)
\!\!\!\!\!\!\!\!\!\!\int\limits_{-(t_0+t_-)/2}^z\!\!\!\!\!\!\!\!
+\,\,\,\theta(t_0+t_-)
\!\!\!\int\limits_{-t_-}^z
\\
&+\,\,\,\theta(t_0-t_-)
\!\!\!\!\!\!\!\!\!\!\int\limits_{-t_-}^{-(t_0+t_-)/2}\!\!\!\!\!\!
-\int\limits_{-t_-}^z\biggr]Q_\Sigma(y)dy
\end{split}
\end{equation*}
and electrons
\begin{equation*}
\label{electronLinearDensityWithScreening}
\begin{split}
N_-(z,t)=\left.N_-\right|_{t=0}(t_+)&+\biggl[\theta(t_0+t_+)
\!\!\int\limits_{z}^{t_+}
+\,\,\,\theta(t_0+t_-)
\!\!\!\!\!\!\!\!\int\limits_{z}^{(t_0+t_+)/2}\!\!\!\!
\\
&+\,\,\,\theta(t_0-t_+)
\!\!\!\!\!\!\!\!\int\limits_{(t_0+t_+)/2}^{t_+}\!\!\!\!\!
-\,\int\limits_{z}^{t_+}\biggr]Q_\Sigma(y)dy
\end{split}
\end{equation*}
and the longitudinal current
\begin{equation}
\label{currentLinearDensityWithScreening}
\begin{split}
&J_z(z,t)=\left.N_+\right|_{t=0}(-t_-)+\left.N_-\right|_{t=0}(t_+)\,\,
+\biggl[\theta(t_0+t_+)
\!\!\!\!\!\!\!\!\!\!\int\limits_{-(t_0+t_-)/2}^{t_+}\!\!\!\!\!\!
\\
&+\,\,\,\theta(t_0+t_-)
\!\!\!\!\!\!\!\!\int\limits_{-t_-}^{(t_0+t_+)/2}\!\!\!\!
+\,\,\,\theta(t_0-t_+)
\!\!\!\!\!\!\!\!\int\limits_{(t_0+t_+)/2}^{t_+}\!\!\!\!\!
+\,\,\,\theta(t_0-t_-)
\!\!\!\!\!\!\!\!\!\!\int\limits_{-t_-}^{-(t_0+t_-)/2}\!\!\!\!\!\!
-\,\int\limits_{-t_-}^{t_+}\biggr]Q_\Sigma(y)dy.
\end{split}
\end{equation}
We will not need the expression for the linear charge density~$N_\rho$. The expressions for $N_+$ and $N_-$ derived from Eqs.~\eqref{smallJzSolution}--\eqref{NplusMinusViaJzAndRho} after the substitution $t=t_{scr}$ should be taken as $\left.N_+\right|_{t=0}(z)$ and $\left.N_-\right|_{t=0}(z)$.

The total number of produced electron-positron pairs is equal to half the value obtained by integrating Eq.~\eqref{currentLinearDensityWithScreening} over the longitudinal coordinate~$z$:
\begin{equation*}
\label{totalPairNumberWithScreening}
N_\Sigma(t)=\left.N_\Sigma\right|_{t=0}+N^{sat}_\Sigma,
\end{equation*}
where
\begin{equation}
\label{Nsaturation}
N^{sat}_\Sigma=\biggl[\,\,t
\!\!\!\!\int\limits_{-t_0-t}^{t_0+t}\!\!
+\!\int\limits_{-t_0-t}^{-t_0}\!\!\!\!(t_0+y)
\,\,+\!\int\limits_{t_0}^{t_0+t}\!\!(t_0-y)
\biggr]Q_\Sigma(y)dy
\end{equation}
is the total number of electron-positron pairs produced after the onset of longitudinal electric field screening. In this formula, the number of pairs produced before the onset of screening acts the number of electron-positron pairs at the initial time, i.e., $\left.N_\Sigma\right|_{t=0}=N^{scr}_\Sigma$ (see Eq.~\eqref{finalTotalParticleNumberBeforeScreening}). It remains to use Eqs.~\eqref{screeningCurrent} and \eqref{Qext} and to write the function
\begin{equation*}
\label{QsumExpr}
Q_\Sigma(z)=\frac{\gamma_{ext}}{2\alpha\tau_{st}},
\end{equation*}
where $\gamma_{ext}$ and $\tau_{st}$ are considered as functions of the longitudinal coordinate~$z$. Recalling the dependence of the particle acceleration time on the stationary Lorentz factor and the external longitudinal electric field~$E^{ext}_z$, we ultimately find
\begin{equation}
\label{finalQsumExpr}
Q_\Sigma(z)=\frac{E_z^{ext}}{2\alpha}.
\end{equation}
At $E_z^{ext}\sim10^{-6}$, we have $Q_\Sigma(z)\sim10^{-4}$.

\section{CALCULATION FOR A DIPOLE MAGNETIC FIELD}

The subsequent calculation of the specific number of electron-positron pairs requires knowing the dependence of the external longitudinal electric field $E^{ext}_z$ on the parameter $z$ along a magnetic field line. This dependence is completely determined by the production point of a primary electron-positron pair in a neutron star magnetosphere. However, the following considerations can be used for qualitative estimates. Let the primary electron-positron pair be produced at a distance $r$ from the stellar center. The total length of the field line fixed by the pair production point will then also be of the order of~$r$. In this case, the length of its part located above the neutron star surface is equal in order of magnitude to $r-R_S$. For simplicity, the produced secondary particles can be assumed to move along the radial coordinate. We see from the expressions for Deutsch's electromagnetic field \citep{Deutsch1955} that the electric field changes along the radial coordinate as a power law. For the subsequent estimates, we will assume that
\begin{equation}
\label{powerEext}
E^{ext}(r)\sim\frac{k_\Omega m R_S^2}{r^4},
\end{equation}
where $m$ is the magnitude of the neutron star magnetic moment, $k_\Omega=1/R_L$ is the wave number corresponding to the angular frequency $\Omega$ of neutron star rotation, and $R_L=c/\Omega$ is the light cylinder radius written in dimensionless units.

Let us first calculate the total number of produced electron-positron pairs for the general model power dependence
\begin{equation}
\label{modelPowerDep}
Q_\Sigma(z)=A_0(z+z_0)^{-u}.
\end{equation}
We used the fact that under the assumptions described above, $r=z+z_0$, where $z_0$ is the distance from the neutron star center at which the primary electron-positron pair is produced from which the plasma tube is subsequently formed. We will also assume that $u>2$. The substitution of dependence \eqref{modelPowerDep} into Eq.~\eqref{Nsaturation} and subsequent integration then yield the expression
\begin{equation}
\label{pairNumberForPowerDepSat}
\begin{split}
N^{sat}_\Sigma=\frac{A_0}{(u-1)(u-2)}
\Bigl[(z_0-t_0-t)^{-u+2}+(z_0+t_0+t)^{-u+2}\\
-(z_0-t_0)^{-u+2}-(z_0+t_0)^{-u+2}\Bigr].
\end{split}
\end{equation}

We will use the fact that the switch-on time of longitudinal electric field screening (in dimensionless units) is small compared to the neutron star radius. Consider the case where the distance from the neutron star center at which the primary electron-positron pair is produced is large compared to the stellar radius. We can then write the inequalities
\begin{equation*}
\label{modelConditions}
t_0\ll R_S\ll z_0.
\end{equation*}
If, in this case, we take
\begin{equation*}
\label{tCond}
t=z_0-t_0-R_S,
\end{equation*}
then it follows from Eq.~\eqref{pairNumberForPowerDepSat} that
\begin{equation*}
\label{simplifiedPairNumberForPowerDep}
N^{sat}_\Sigma=\frac{A_0}{(u-1)(u-2)}R_S^{-u+2}.
\end{equation*}
The latter quantity is the number of electron-positron pairs produced in the regime of saturation by the instant the plasma tube reaches the neutron star surface. Let us now return to Eq.~\eqref{finalQsumExpr}. If we take into account Eq.~\eqref{powerEext}, then $A_0=k_\Omega mR_S^2/2\alpha$ and $u=4$ correspond to it. Then,
\begin{equation*}
\label{NpairSat}
N_{\Sigma}^{sat}=\frac{k_\Omega m}{12\,\alpha}
\end{equation*}
is the characteristic number of electron-positron pairs produced in the plasma tube after the onset of longitudinal electric field screening, so that the plasma is generated in the regime of saturation of the total source. It is easy to see that $N_{\Sigma}^{sat}$ depends only on the dipole magnetic moment of the neutron star and on the angular frequency of its rotation. If the dipole magnetic moment $m$ of the neutron star is expressed in terms of its radius $R_S$ and the magnetic field strength $B_0$ at its magnetic pole, then
\begin{equation}
\label{NpairSatViaBsurf}
N_{\Sigma}^{sat}=\frac{B_0}{24\alpha}\frac{R_S^3}{R_L}.
\end{equation}
For $R_S\approx2.6\times10^{16}$ ($10$~km in dimensional units), $R_L=10^4R_S$, and $B_0\sim0.01$, we have $N_{\Sigma}^{sat}\sim10^{28}$. If, alternatively, the magnetic field strength at the magnetic equator of the neutron star is meant by~$B_0$, then the right-hand side of Eq.~\eqref{NpairSatViaBsurf} should formally be doubled.

Thus, we can write the total number of electron-positron pairs produced in the plasma tube from the instant of primary pair production by a photon from the external cosmic gamma-ray background to the instant the plasma tube reaches the neutron star surface,
\begin{equation*}
\label{Nsigma}
N_\Sigma=N^{scr}_\Sigma+N^{sat}_\Sigma,
\end{equation*}
where $N^{scr}_\Sigma$ and $N^{sat}_\Sigma$ are defined by Eqs.~\eqref{finalTotalParticleNumberBeforeScreening} and~\eqref{NpairSatViaBsurf}, respectively. Given the inequality
\begin{equation*}
\label{NscrToNsatRatio}
N^{scr}_\Sigma\ll N^{sat}_\Sigma,
\end{equation*}
we can ultimately write
\begin{equation*}
\label{NsigmaFinal}
N_\Sigma\approx N^{sat}_\Sigma.
\end{equation*}

It remains to verify that the effective source $Q^{eff}$ can be used over the entire plasma generation time. This requires that the condition $N_\tau\gg1$, where $\tau=\tau_{st}$ is the delay time, be met. This inequality holds if the particle Lorentz factor exceeds $10^3$ \citep{IstominSobyanin2011a}. The longitudinal current can be estimated from the formula
\begin{equation}
\label{JparallelEstimation}
J_z\lesssim2\frac{N^{sat}_\Sigma}{R_S}.
\end{equation}
In order of magnitude, $J_z\sim10^{11}-10^{12}$, which corresponds to $J_z/J_0\sim10^2-10^3$. Consequently, the longitudinal electric field $E_z$ decreases by three orders of magnitude compared to the initial external electric field, which corresponds to $E^{f}_z\sim10^{-9}$ for $E^{ext}_z\sim10^{-6}$. It follows from Eq.~\eqref{gamma0approximation} that the Lorentz factor decreases by only one order of magnitude compared to the initial one, $\gamma^f_0\approx0.1\gamma_{ext}$, and is $\gamma^f_{0}\sim10^7$ for $\gamma_{ext}\sim10^8$.

Note the interesting relation
\begin{equation*}
\label{NsigmaToNgjRatio}
\frac{N_\Sigma}{n_{GJ}V_S}
\approx\frac{1}{16},
\end{equation*}
where $V_S=4\pi R_S^3/3$ is the volume of the neutron star and $n_{GJ}$ is the Goldreich-Julian density near its surface. It shows that the initial filling of the magnetosphere is very efficient. Only $10-100$ absorbed photons from the external cosmic gamma-ray background are enough to fill the inner neutron star magnetosphere. The lightnings forming in this case contain a number of particles comparable to that in the inner magnetospheric regions completely filled with plasma. If, alternatively, we consider the entire magnetosphere up to the light cylinder~$R_L$, then the number of particles in it is equal in order of magnitude to $3n_{GJ}V_S\ln(R_L/R_S)$. At~$R_L/R_S\sim10^3-10^4$, about $300-500$ absorbed photons are needed to fill the entire magnetosphere.

\section{THE PLASMA TUBE RADIUS AND MULTIPLICITY}

Let us estimate the multiplicity
\begin{equation}
\label{multiplicity}
\lambda_m=\frac{n}{n_{GJ}},
\end{equation}
where $n$ is the number density of electron-positron pairs in the plasma tube and $n_{GJ}$ is the Goldreich-Julian density equal in dimensionless units to
\begin{equation*}
\label{GJdimentionlessDensity}
n_{GJ}=\frac{|\mathbf{\Omega}\cdot\mathbf{B}|}{2\pi\alpha}.
\end{equation*}
For our estimates, we can neglect the dependence on the angle between the vectors $\mathbf{\Omega}$ and $\mathbf{B}$ present in this formula and write
\begin{equation}
\label{GJorder}
n_{GJ}\approx\frac{1}{2\pi\alpha}\frac{B}{R_L}.
\end{equation}
We know typical values of the longitudinal current~$J_z$. To estimate the number density~$n$, the radius of the forming tube should be known.

It is easy to estimate the radius of the plasma tube if the characteristic velocity of its boundary is known. This velocity is determined by two factors. The first factor is the electric drift of particles at the tube boundary in crossed electric and magnetic fields (see Eq.~\eqref{electronPositronVelocity}). In order of magnitude, the electric drift velocity can reach $E/B\sim10^{-4}$ (in units of the speed of light). The second factor is the motion of the tube boundary due to the finite photon mean free path~$l_f$. The point is that the photon emitted by a particle almost along the tangent to the magnetic field line produces an electron-positron pair on a different field line shifted relative to the initial field line by the distance
\begin{equation*}
\label{orthogonalDisplacement}
l_\perp=\frac{l^2_f}{2\rho}
\end{equation*}
due to the curvature of the field line itself. This distance corresponds to the orthogonal velocity of the boundary
\begin{equation*}
\label{orthogonalBoundaryVelocity}
v_\perp=\frac{l_f}{2\rho}.
\end{equation*}
For our estimates, as $l_f$ we can take $l_{\max}=\rho\chi_{\max}$, where $\chi_{\max}$ is related to the minimum longitudinal Lorentz factor of the produced particles as $\gamma_{\min}=1/\chi_{\max}$ \citep{IstominSobyanin2007}. We will then obtain the following qualitative estimate:
\begin{equation}
\label{orthogonalBoundaryVelocityQualitativeEstimation}
v_\perp=\frac{1}{2\gamma_{\min}}.
\end{equation}
For $\gamma_{\min}\sim10^2$, we obtain $v_\perp\sim10^{-2}$ from this estimate. We see that the finite mean free path can increase considerably the velocity of the boundary compared to what the ordinary electric drift velocity can give. The time dependence of the tube radius will be given by the formula
\begin{equation*}
\label{pipeRadius}
R=v_\perp t,
\end{equation*}
where, for simplicity, the velocity of the boundary is assumed to be constant. For this estimate, the tube radius on time scales $t\sim R_S$ will reach
\begin{equation}
\label{radiusEstimation}
R\lesssim100\text{ m}.
\end{equation}

To estimate the multiplicity, we will now take the electric current estimate from Eq.~\eqref{JparallelEstimation}, in which a tube length equal to $R_S$ was taken. The corresponding time is $t=R_S/2$. The number density of electron-positron pairs is then
\begin{equation*}
\label{pairVolumeDensity}
n=\frac{2}{v^2_\perp}\frac{J_z}{\pi R_S^2}.
\end{equation*}
Using Eqs.~\eqref{NpairSatViaBsurf}--\eqref{GJorder}, we ultimately obtain
\begin{equation*}
\label{finalMultiplicity}
\lambda_m\approx\frac{2}{3v^2_\perp}.
\end{equation*}
Since we took $B_0$ as the magnetic field on the magnetic equator of the neutron star, Eq.~\eqref{NpairSatViaBsurf} was modified in accordance with the remark made after it. We see that to determine typical values of the multiplicity, it will suffice to know only the expansion velocity of the forming plasma tube. The lower multiplicity limit can be derived using Eq.~\eqref{orthogonalBoundaryVelocityQualitativeEstimation} after the substitution $\gamma_{\min}\sim100$. The upper limit can be obtained if a typical drift velocity $E/B\sim10^{-4}$ is taken as the velocity of the boundary. Thus, we have
\begin{equation*}
\label{MultiplicityBoundaries}
10^4<\lambda_m<10^8.
\end{equation*}
Obviously, the upper limit is grossly overestimated, because the finiteness of the photon mean free path is disregarded, and the actual multiplicity is closer to the lower limit.

\section{CONCLUSIONS}

The absorption of a high-energy photon from the external cosmic gamma-ray background in the inner neutron star magnetosphere triggers the generation of a secondary electron-positron plasma and gives rise to a lightning---a lengthening and simultaneously expanding plasma tube. It propagates along magnetic field lines with a velocity close to the speed of light. The orthogonal velocity of the tube side boundary can reach $0.01$ of the speed of light. Such a velocity is provided by the pair production on magnetic field lines shifted relative to those along which the photon-emitting particles move.
 
The shift is determined by the finite photon mean free path and the curvature of the magnetic field lines. The existence of a strong longitudinal electric field provides the acceleration of particles and the emission of curvature and synchrotron photons by them, which, in turn, also produce electron-positron pairs. Initially, the plasma is produced in the external vacuum electromagnetic field, while there is no screening of this field. After a time $t_{scr}\sim10^{-7}-10^{-6}$~s, the electric current in the lightning becomes equal to the screening current $J_0\sim10^{11}$~A. Note that the screening current is comparable to the characteristic electric current flowing through the polar region of the magnetosphere of an ordinary radio pulsar. About~$10^{22}$ electron-positron pairs are produced by the instant the screening current in the lightning is reached. Subsequently, the dynamical screening of the longitudinal electric field through the growing electron-positron generation rate begins. It is provided not by static charge separation but by electric current growth in the lightning. We emphasize that the production of an electron-positron plasma in a lightning is essentially nonstationary, while the dynamical electric field screening in the stationary case is impossible.

The plasma is mainly produced in the regime of screening, when the longitudinal electric field in the lightning and the source of electron-positron pairs are determined self-consistently from the following condition: the longitudinal electric field depends on the electron-positron pair generation rate, which itself depends on the electric field in the lighting. The time it takes for the lightning to reach the neutron star surface may be considered as its lifetime. Subsequently, the screening mechanism can change due to the influence of the neutron star surface. The plasma production can even cease, for example, due to the onset of static electric field screening. The electric current flowing in the lightning when it reaches the neutron star surface is higher than the screening current by two or three orders of magnitude. Although the longitudinal electric field by the end of lightning formation decreases by three orders of magnitude compared to the initial vacuum electric field, the particle Lorentz factor decreases by only an order of magnitude and reaches $10^6-10^7$. This value is much larger than the characteristic Lorentz factor of the secondary-plasma particles in the magnetosphere of an ordinary pulsar.

If the production of an electron-positron plasma is assumed to be efficient in a magnetic field higher than $10^8$~G, then the distance of $(10-100)R_S$ in exceeding which the plasma generation begins to be suppressed corresponds to a neutron star surface magnetic field of $10^{12}-10^{14}$~G. Consequently, the lightning length can reach $1000$~km. The lightning radius is $100$~m. Interestingly, the lightning radius is comparable to the radius of the polar regions in a neutron star magnetosphere formed by open field lines. The number of electron-positron pairs produced in the lightning in its lifetime reaches~$10^{28}$. This corresponds to the constraints on the multiplicity $\lambda_m>10^4$, which is comparable to or even larger than that in the polar regions of ordinary pulsars.

We see that the parameters of the generated plasma are comparable to those of the electron-positron plasma in the polar regions of ordinary radio pulsars. Indeed, the multiplicity is comparable to that in the magnetosphere of a radio pulsar. The particle Lorentz factor exceeds considerably that of the secondary plasma in pulsars and magnetars. The curvature of the magnetic field lines is comparable to that in the polar cap of a radio pulsar if the lightning is formed in the region of open field lines or exceeds it considerably if the lightning is formed in the region of the magnetic equator. In addition, the lightning width is comparable to the polar cap width of a radio pulsar. In view of the existing analogy, these factors suggest that the radio emission can be generated by individual lightnings.

If we consider a neutron star located at a distance of the order of the distance to a typical pulsar, then we can assume the possibility of observing radio bursts from lightnings at existing radio astronomy observatories. Moreover, one may expect the radio flux density in a lightning to exceed the flux density from ordinary radio pulsars due to larger particle Lorentz factor, multiplicity, and plasma density. One of the factors determining such parameters is the following difference between the plasma generation mechanisms in a lightning and in the polar cap of a radio pulsar. If we consider an ordinary radio pulsar with a relatively weak (compared to the critical one) magnetic field, then, apart from the primary particles accelerated in the polar gap, there are at least two more generations of secondary particles: the first generation of secondary particles produced by the curvature photons from primary particles and the second generation of particles produced by the synchrotron photons emitted through the transition of first-generation particles to the zeroth Landau level. However, a nearly vacuum electric field is available only in the polar gap up to distances of about $100$~m above the neutron star surface. Above the gap, the electric field is screened. In contrast, in the case of a lightning, even despite the dynamical screening, we have a fairly significant longitudinal electric field that continuously accelerates the newly produced particles. These particles, whatever generation they formally belong to, immediately begin to play the role of primary particles after their acceleration. Obviously, the electric field in a lightning cannot strictly become zero. If this occurred, then the electric field screening would immediately disappear and the field itself would become close to Deutsch's vacuum field. Consequently, the electric field is always nonzero and the plasma is generated continuously as long as the lightning exists.

Stationary electron-positron plasma outflows from the magnetosphere are responsible for the radio emission from pulsars, but the plasma generation in a lightning is nonstationary. Consequently, the radio emission from lightnings, if it can be observed, must be connected with the observational manifestations of nonstationary radio sources associated with neutron stars. The formation time of radio emission is limited by the lightning lifetime. If the width of a radio burst is assumed to be comparable in order of magnitude to the ratio of the lightning length to the speed of light, then we have a burst width of $3.3$~ms for a lightning length of $1000$~km. Radio bursts of comparable width are observed from rotating radio transients. For example, the duration of a single burst from RRAT J1819--1458 is $3$~ms in observations at $1.4$~GHz \citep{McLaughlinEtal2006}. Here, we disregard the pulse broadening due to scattering, which is important in RRAT observations at lower radio frequencies \citep{Shitov2009}. As regards the flux density, it reaches $10$~Jy at 1.4~GHz \citep{KeaneEtal2010}, exceeding considerably that for typical radio pulsars.

Thus, we can assume that the radio bursts from lightnings can be observed as those from RRATs. Investigation of the observational manifestations of lightnings in a neutron star magnetosphere and their possible connection with radio bursts from RRATs is beyond the scope of this paper and requires a separate consideration.

\begin{acknowledgments}
This work was supported in part by the Russian Foundation for Basic Research (project no.~11-02-01021-a).
\end{acknowledgments}
\newpage
\bibliography{JETP2}
\begin{flushright}
\textit{Translated by V. Astakhov}
\end{flushright}
\newpage
\begin{figure}[h]
\centering
\begin{overpic}[angle=-90]{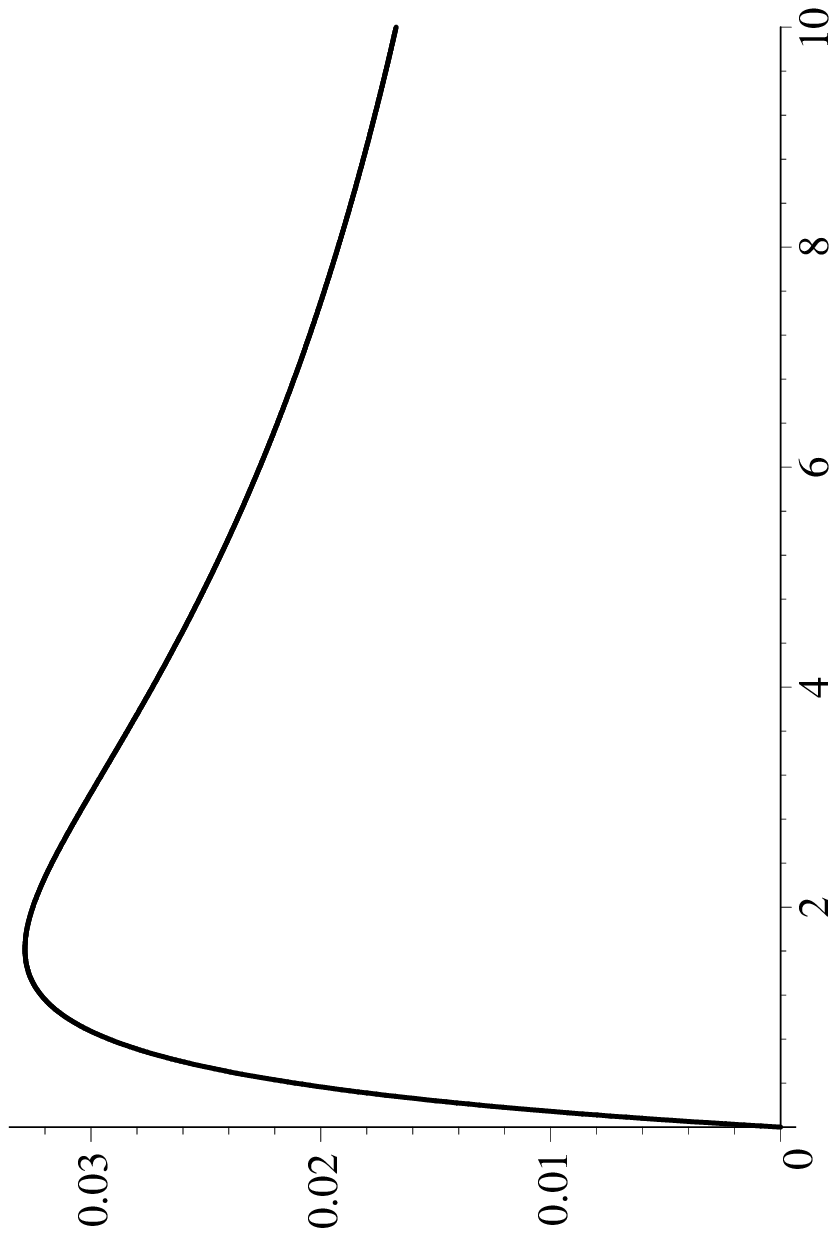}
\put(47,-5){$J_z/J_0$}
\put(-16,35){\large$\frac{\gamma_0-\gamma^{ap}_0}{\gamma_0}$}
\end{overpic}
\newline\caption{Accuracy of the approximation $\gamma_0^{ap}$ of the Lorentz factor $\gamma_0$ versus ratio of the electric current $J_z$ flowing in the lightning to the screening current~$J_0$.}
\end{figure}
\end{document}